\documentclass{elsart}

\usepackage{epsfig}

\newcommand{\be}{\begin{equation}}
\newcommand{\ee}{\end{equation}}
\newcommand{\bea}{\begin{eqnarray}}
\newcommand{\eea}{\end{eqnarray}}
\newcommand{\pT}{p_\perp}
\newcommand{\kTaveSq}{\mbox{$\left\langle k_\perp^2\right\rangle$}}

\newcommand{\Pythia}{\textsc{Pythia}}
\newcommand{\Jetset}{\textsc{Jetset}}
\newcommand{\Fritiof}{\textsc{Fritiof}}

\newcommand{\UNIT}[1]{\mbox{$\,{\rm #1}$}}

\newcommand{\GeV}{\UNIT{GeV}}
\newcommand{\GeVc}{\UNIT{GeV/c}}

\newcommand{\fmc}{\UNIT{fm/c}}
\newcommand{\proz}{\UNIT{\%}}
\newcommand{\SqrtS}[1]{\mbox{$\sqrt s=#1\GeV$}}

\begin{document}

\begin{frontmatter}
\title{Jet quenching  by (pre--)hadronic final state interactions at
RHIC}
\author[unig]{K.~Gallmeister\corauthref{cor1}}
\ead{Kai.Gallmeister@theo.physik.uni-giessen.de}
\corauth[cor1]{corresponding author}
\author[unig]{W.~Cassing}
\address[unig]{Institut f\"ur Theoretische Physik, %
  Universit\"at Giessen, %
  Heinrich--Buff--Ring 16, %
  D--35392 Giessen, %
  Germany}

\begin{abstract}
  Within a hadron-string dynamical transport approach (HSD) we investigate
  the attenuation of high transverse momentum ($\pT$) hadrons as well as the
  suppression of 'near-side' and 'far-side' jets in $Au+Au$
  collisions at invariant energies $\sqrt{s}$ = 200 GeV and $\sqrt{s}$ =
  62.4   GeV in comparison to the data available from
  the Relativistic Heavy-Ion Collider (RHIC).  From our transport studies we find that
   a significant part of the high $\pT$ hadron attenuation seen
  experimentally can be attributed to inelastic interactions of 'leading'
  pre-hadrons with the dense hadronic environment.
  In addition, we also show   results of 'near-side' and
  'far-side' angular correlations of high $\pT$ particles from  Au+Au collisions
  at $\sqrt{s}$ = 200 GeV and $\sqrt{s}$ = 62.4 GeV within this (pre-)hadronic
  attenuation scenario.
  It turns out that the 'near-side' correlations are unaltered -- in
  accordance with experiment --
  whereas the 'far-side' correlations are suppressed by up to $\sim$ 60\% in central
  collisions. Since a much larger suppression is observed experimentally
  for these reactions in central reactions we conclude that
 there should be strong additional (and earlier) partonic
  interactions in the dense and possibly colored medium created in Au+Au collisions at RHIC.
\end{abstract}


\end{frontmatter}

%\newpage

\section{Introduction}
The phase transition from partonic degrees of freedom (quarks and
gluons) to interacting hadrons  is a central topic of modern
high--energy physics. In order to understand the dynamics and
relevant scales of this transition -- as well the dynamical
formation of hadrons -- laboratory experiments are presently
performed with ultra--relativistic nucleus--nucleus collisions.
 Estimates based on the Bjorken formula \cite{bjorken}
and/or calculations within  hadron-string-dynamical transport
approaches \cite{CassKo,URQMD1,URQMD2} indicate that the critical
energy density for the formation of a quark--gluon plasma (QGP) of
0.7 to 1 GeV/fm$^3$ \cite{Karsch} should by far be exceeded in the
initial phase of central Au+Au collisions for a couple of $\fmc$
at top SPS and RHIC energies
\cite{QM01}. However, the task still remains to unambiguously
identify the formation and properties of this new phase.

Whereas there are a couple of tentative signatures for an early
partonic phase such as the high transverse pressure seen
experimentally \cite{survey,Miklosrev} as well as the large
elliptic flow of hadrons at moderate and high momenta
\cite{survey,STARv2}, a unique 'smoking gun' could not be
identified so far. In order to find representative signatures one
has to look for early 'hard' probes that dominantly test the first
few fm/c of the relativistic heavy-ion collision. Hence, the
measurements of 'hard' jets should offer a direct access to the
dynamics in the early stage when the (possibly deconfined) matter
is very dense. In fact, the PHENIX \cite{PHENIX1} and STAR
\cite{STAR1} collaborations have reported a large relative
suppression of hadron spectra for transverse momenta $\pT$ above
$\sim3-4\GeVc$ which might point towards the creation of a QGP,
since this suppression is not observed in d+Au interactions at the
same bombarding energy per nucleon \cite{BRAHMS,PHENIX2,STAR2} and
thus has to be attributed to final state interactions.
Furthermore, an almost complete suppression of 'far-side' jets has
been reported for central Au+Au collisions \cite{PHENIXnew} which
is also not seen in d+Au reactions \cite{dAu}.

These observations -- at first sight -- appear to be compatible
with the picture of partonic energy loss in a dense colored medium
\cite{WA1,WA2,Wang,Baier,vitev,R1,R2,R3,R4,R5,R6,Wangxxx,Dainese,Eskola},
however, it is not clear presently to which extent these
suppression phenomena might be due to ordinary hadronic
final-state-interactions (FSI) \cite{Kai} since (in-)elastic
collisions of (pre-)hadronic high momentum states with the bulk of
hadrons in the late fireball might also contribute significantly
to the attenuation of the high transverse momentum hadrons at RHIC
\cite{Kahana,Humanic,CGG}. Hence, the  hadronic attenuation has to
be addressed in more detail before conclusions on perturbative QCD
effects in a deconfined QGP phase on the materializing jets can be
drawn. In fact, as shown in Refs. \cite{CGG,CGG2} a major fraction
of the high $\pT$ hadron attenuation in Au+Au collisions at
$\sqrt{s}$ = 200 GeV can be described by the (dominantly)
inelastic interactions of 'leading' pre-hadrons.

In order to specify the latter findings in Refs. \cite{CGG,CGG2}
we briefly recall the concept of 'leading' (pre-hadronic) and
'secondary' (ordinary) hadrons in the string-hadron transport
approach: In a high energy nucleon-nucleon collision two (or more)
color-neutral strings are assumed to be formed; the string ends
are defined by the space-time coordinates of the constituents
which are denoted as 'leading' quarks. The latter are initially
colored but assumed to pick up almost instantly a quark (or
antiqiuark) from the vacuum to build up color neutral 'leading'
pre-hadrons \cite{Kopel4}. These pre-hadronic states are
distinguished from 'secondary' hadrons that arise from the further
fragmentation of the strings and consist solely of quarks and
antiquarks (diquarks and antidiquarks) created from the vacuum at
some later time. The total time that is needed for the
fragmentation of the strings and for the hadronization of the
fragments is denoted as {\it formation time} and taken as $\tau_f
\approx 0.8$ fm in the HSD transport approach. Due to time
dilatation the formation time $t_f$ in any reference frame is then
proportional to the Lorentz $\gamma$-factor,
\begin{equation}
t_f = \gamma \tau_f. \end{equation} We stress that the formation
time is a sensible parameter and has been fixed in \cite{Ehehalt}
by the rapidity distribution of charged particles at SPS energies
in the range 0.6--0.8 fm/c. Though $\tau_f$ is no longer a free
parameter we will perform also calculations for the jet
suppression in Section 4 for different $\tau_f$ ranging from 0.2
to 0.5 fm/c. In principle, other scenarios for modeling the string
fragmentation might also be investigated, e.g. formation times
depending explicitly on the hadron species or those extracted from
the LUND fragmentation scheme (cf. Ref. \cite{Falter2}); however,
we here constrain to the simple model introduced in Ref.
\cite{Ehehalt} and delay an investigation of alternative
prescriptions to future studies.

 In the HSD transport approach, furthermore, it is
assumed  that hadrons, whose constituent quarks and antiquarks are
created from the vacuum in the string fragmentation, do {\it not
interact} with the surrounding nuclear medium within their
formation time $t_f$ and thus do {\it not} create early pressure.
On the other hand, for the leading pre-hadrons, i.e.~those
involving quarks (antiquarks) from the struck nucleons,  a reduced
effective cross section $\sigma_{lead}$ is adopted during the
formation time $t_f$ and the full hadronic cross section later on.
The effective cross section $\sigma_{lead}$ is fixed in line with
the additive constituent quark model \cite{CGG} (cf. also Section
3).

As shown in Ref. \cite{CGG} for central Au+Au collisions at
$\sqrt{s}$ = 200 GeV hadrons with transverse momenta larger than
$\sim 6\GeVc$ predominantly stem from the string ends and
therefore can, in principle, interact directly with the reduced
cross section $\sigma_{lead}$ . Moreover, we have to stress that
in the HSD approach the formation of secondary hadrons (e.g. in
Au+Au collisions) is not only controlled by the formation time
$\tau_f$, but also by the energy density in the local rest frame:
hadrons are not allowed to be formed if the energy density is
larger than $1\UNIT{GeV/fm^3}$, which is about the critical energy
density for a phase transition to the QGP in thermal equilibrium
\cite{Karsch}. The interaction of the leading and energetic
(pre-)hadrons with the soft hadronic and bulk matter is thus
explicitly modeled to happen only for local energy densities below
that cut, i.e. pre-hadronic interactions with fully formed hadrons
as well as interactions between formed hadrons are delayed until
the energy density has dropped below $1\UNIT{GeV/fm^3}$ in the
expansion phase! Note, however, that the highest energy densities
are achieved in the center of the 'fireball' and that the
formation of hadrons starts at the surface where substantially
lower energy densities are encountered.

We recall that the interactions of leading (pre--)hadrons are also
important to understand the attenuation of hadrons with high
(longitudinal) momentum in ordinary 'cold' nuclear matter. The
studies in Refs. \cite{Falter2,Falter} (see also \cite{Kopel3})
have shown that the dominant final-state-interactions of the
hadrons with maximum momentum -- as measured by the HERMES
Collaboration \cite{HERMES} -- are compatible with the concepts
described above. These independent investigations impose severe
constraints on hadron attenuation in a medium that might possibly
be attributed to a QGP phase (cf. also Ref. \cite{Wangxxx}).

In this work we concentrate on the transverse momentum dynamics
and especially on the very high momentum tail of the hadron
spectra. In order to describe these high $\pT$ spectra, we use the
\Pythia{} v6.2 event generator \cite{PYTHIA} for nucleon--nucleon
collisions and the HSD transport model \cite{Ehehalt,Geiss,Cass99}
for the space-time localization of the events and further
propagation in time including final-state interactions
(cf. Refs. \cite{CGG,CGG2}).

Our study is organized as follows: We start in Section 2 with an
analysis of high $\pT$ hadron events from $pp$ reactions  --
generated via \Pythia{} v6.2 --  with respect to correlations in
the azimuthal angle $\varphi$ employing different momentum cuts
for the hadrons. Section 3 gives a very brief reminder of the HSD
approach and the actual 'input' to the transport calculations. The
calculated results for jet quenching in Au+Au collisions at
$\sqrt{s}$ = 200 GeV for different centrality classes are
presented in Section 4. The sensitivity of the jet suppression to
the formation time $\tau_f$ will be demonstrated, too. In Section
5 we will show the results of calculations for Au+Au collisions at
$\sqrt{s}$ = 62.4 GeV for different centrality classes since this
system is presently under investigation experimentally. Apart from
a comparison of the high $\pT$ attenuation of charged hadrons with
the data from the PHOBOS collaboration \cite{PHOBOSnew} we will
also present our predictions for the 'near-side' and 'far-side'
angular correlations at this energy. Section 6 summarizes our
present investigations.

\section{Jet correlations in $pp$ reactions}

Before coming to the more subtle issue of jet quenching in Au+Au
collisions at RHIC energies, it is instructive to have a look at
the jet topology in $pp$ reactions. As mentioned before, we employ
the event generator \Pythia{} v6.2 which has been adjusted with
respect to the \kTaveSq{} parameter to properly describe the
transverse hadron spectra at $\sqrt{s}$ = 200 GeV (cf. Fig. 1 in
Ref. \cite{CGG}). As a trigger particle we choose a charged high
$\pT$ hadron in the rapidity interval $|y_{cm}| < 0.5$ with
momentum 4 GeV/c $\leq \pT^{Trig} \leq$ 6 GeV/c as in Ref.
\cite{StarAngCorr}. The correlated particles then are selected in
the interval 2 GeV/c $\leq \pT^{Sec} \leq \pT^{Trig}$ and $|y_{cm}| <
1.0$. The resulting angular correlation
\begin{equation}
\label{ang} C(\Delta \varphi) = \frac{1}{N_{Trig}} \frac{d N}{d
\Delta \varphi}
\end{equation}
with respect to the azimuthal angle of the trigger particle is
displayed in Fig. 1 (upper left part) and shows two distinct peaks
which are attributed to 'near-side' and 'far-side' jet
correlations since the maxima at $\varphi$=0 and $\varphi=\pi$
correspond to back-to-back emission. Furthermore, the comparison
to the data from the STAR Collaboration \cite{StarAngCorr} shows that the
angular correlations measured for $pp$ reactions at $\sqrt{s}$ =
200 GeV are well reproduced by the event generator \Pythia{} v6.2.

%%%%%%%%%%%%%%%%%%%%%%%%%%%%%%%%%%%%%%%%%%%%%%%%%%%%%%%%% Fig.1%%%%%%%%
\begin{figure}[htb!]
  \begin{center}
    \includegraphics[width=13.5cm]{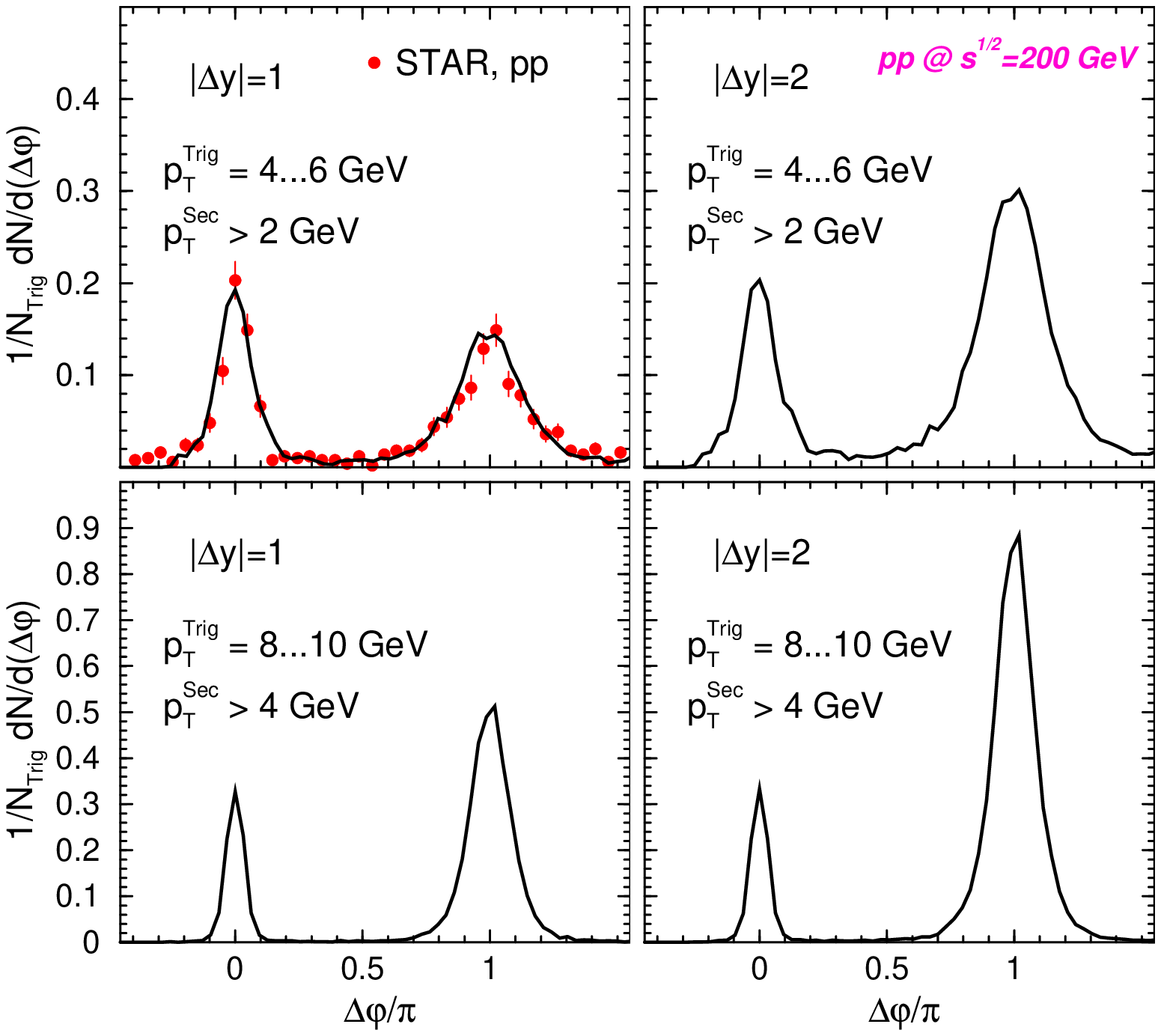}
    \caption{
      The correlation in the azimuthal angle $\varphi$ for $pp$ reactions at
      $\sqrt{s}$ = 200 GeV for different cuts on the trigger particle and
      associated particles from the event generator \Pythia{} v6.2.
      The data in the upper left part are taken from Ref. \cite{StarAngCorr}.}
    \label{fig1}
  \end{center}
\end{figure}
%%%%%%%%%%%%%%%%%%%%%%%%%%%%%%%%%%%%%%%%%%%%%%%%%%%%%%%%%%%%%%%%%%%%%%%

The structure and relative height of the two peaks changes when
selecting momenta 8 GeV/c $\leq \pT^{Trig} \leq$ 10 GeV/c  for the
trigger particle and 4 GeV/c $\leq \pT^{Sec} \leq \pT^{Trig}$ for
the other charged hadrons (cf. Fig. 1, lower left part). In case
of the higher momentum trigger particle the angular distribution
of associated hadrons (from jet fragmentation) becomes narrower;
the 'far-side' peak gets larger since the number of associated
hadrons in the peaks are lower than for the cuts in the upper left
part of Fig. 1. The asymmetry in the height of the peaks is due to
the fact that the 'trigger particle' is not counted in the
'near-side' angular correlation. Apart from the sensitivity to the
momentum cuts demonstrated above the height of the peaks also
depends on the rapidity cut for the associated particles. When
gating on a larger rapidity interval $|\Delta y| = 2$ (r.h.s. of Fig. 1)
the 'near-side' peak roughly stays the same whereas the 'far-side' peak increases
by almost a factor of two.

 A more detailed
information is provided in the upper part of Table 1 where the probability of
finding 0, 1, $\cdots$, 3 further hadrons in the 'near-side' and
'far-side' peak is given for 4 GeV/c $\leq \pT^{Trig} \leq$ 6 GeV/c and
 2 GeV/c $\leq \pT^{Sec} \leq \pT^{Trig}$ and $|y_{cm}| <
1.0$. These numbers demonstrate, that in less than 10\% of all
events a further particle is found in the given rapidity and $\pT$
interval and that the conditional probability to find any further
particle roughly decreases by an order of magnitude. These low
conditional probabilities should be kept in mind when comparing
azimuthal angular correlations from Au+Au collisions (cf. Sections
4 and 5). We note additionally that 'jet structures' only show up
after averaging over a large event sample.

%%%%%%%%%%%%%%%%%%%%%%%%%%%%%%%%%%%%%% Table 1 %%%%%%%%%%%%%%%%%%%%%%%%%
\begin{table}[htb!]
  \begin{center}
\begin{tabular}{|c||cccc|}
\hline
\#particles &
\multicolumn{4}{c|}{\#particles far side}\\
near side      &    0      &      1     &      2     &     3       \\
\hline
\multicolumn{5}{|c|}{pp @ 200 GeV}\\
\hline
    0 &  0.89 &  0.047 &  0.0060 &  0.00062 \\
    1 &  0.046 &  0.0078 &  0.0020 &  0.00031 \\
    2 &  0.0026 &  0.0010 &  0.00049 &  0.00012 \\
    3 &  0.00030 &  0.00013 &  0.000055 &  0.000017 \\
\hline
\multicolumn{5}{|c|}{central Au+Au @ 200 GeV}\\
\hline
    0 &  0.93 &  0.024 &  0.0024 &  0.00034 \\
    1 &  0.040 &  0.0033 &  0.00043 &  0.000023 \\
    2 &  0.0031 &  0.00041 &  0.00016 &  0.000021 \\
    3 &  0.00040 &  0.00013 &  0.0000065 &  0 \\
\hline
\end{tabular}
\caption{The conditional probability to find in addition to the
trigger particle 0,1,\dots3 particles in the 'far-side' and 0,1,\dots3
particles in the 'near-side' peak in $pp$ (upper part) and central Au+Au collisions
(lower part) at \SqrtS{200}.
The trigger conditions are: 4 GeV/c $\leq \pT^{Trig} \leq$ 6 GeV/c and
 2 GeV/c $\leq \pT^{Sec} \leq \pT^{Trig}$ and $|y_{cm}| <
1.0$.
}
\label{tab1}
  \end{center}
\end{table}
%%%%%%%%%%%%%%%%%%%%%%%%%%%%%%%%%%%%%%%%%%%%%%%%%%%%%%%%%%%%%%%%%%%%%%%%
%%%%%%%%%%%%%%%%%%%%%%%%%%%%%% Fig. 2 %%%%%%%%%%%%%%%%%%%%%%%%%%%%%%%%%%%
\begin{figure}[htb!]
  \begin{center}
    \includegraphics[angle=-90,width=10.0cm]{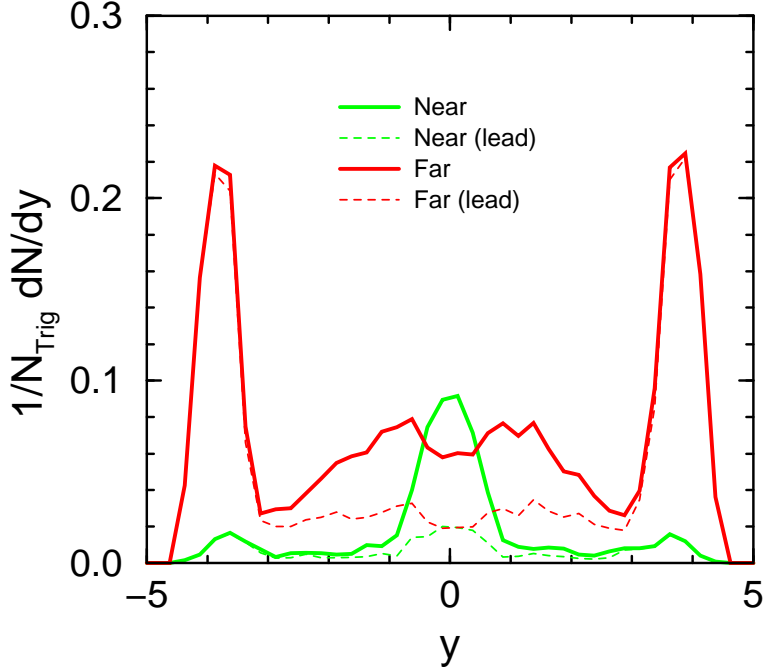}
    \caption{
      The rapidity distribution of correlated hadrons when gating on a high $\pT$
      hadron in the interval 4 $\leq \pT^{Trig} \leq $ 6 GeV/c for $|y| <$ 0.5
      and  2 $\leq \pT^{Sec} \leq \pT^{Trig}$. The 'near-side' contribution is also
      peaked at midrapidity (lower solid line, color: green), however, shows only a small
      content of 'leading' particles (lower dashed line, color: green). On the other hand,
      the rapidity distribution of correlated hadrons in the 'far-side' peak
      (upper solid line, color: red) is widespread with maxima around
      $|y| \approx$ 4 dominantly due to leading hadrons
      (upper dashed line, color: red).  }
    \label{fig2}
  \end{center}
\end{figure}
%%%%%%%%%%%%%%%%%%%%%%%%%%%%%%%%%%%%%%%%%%%%%%%%%%%%%%%%%%%%%%%%%%%%%%%%%

It is, furthermore, instructive to have a look at the rapidity
distribution of charged particles when gating on a trigger hadron
in the rapidity interval $|y_{cm}| < 0.5$ with momentum 4 GeV/c
$\leq \pT^{Trig} \leq$ 6 GeV/c. The correlated particles in the
'far-side' peak --  selected in the interval 2 GeV/c $\leq
\pT^{Sec} \leq \pT^{Trig}$ -- then show a rather wide distribution
in rapidity as seen from Fig. 2. Here, the upper solid line
corresponds to the correlated charged hadrons in the 'far-side'
peak (color: red) whereas the lower solid line corresponds to the
correlated charged hadrons in the 'near-side' peak (color: green)
which is localized in rapidity around the trigger particle. The
dashed lines give the rapidity distribution for the correlated
charged hadrons in the 'far-side' (upper dashed line, color: red)
and 'near-side' peak (upper dashed line, color: red) which
correspond to leading particles, i.e. where the hadron carries at
least one of the constituent quarks of the initial protons. Note
that in $pp$ collisions two strings are formed with in total four
leading particles. For the momentum cuts adopted the leading
particles show up dominantly at $|y| \approx$ 4, i.e. in the
rapidity regime of the projectile/target remnants.

Consequently, the hadrons produced and/or formed by the
fragmentation of jets do not occupy a narrow rapidity window close
to the trigger hadron but are widespread in rapidity at least in
the 'far-side' peak (with even a small dip at midrapidity). This
wide distribution also explains why the probability of observing a
correlated hadron together with a trigger hadron in roughly the
same rapidity interval $|\Delta y| \approx 1$ and high momentum is
rather low (cf. upper part of Table 1).

\section{Ingredients of the transport model}

We employ the HSD transport model \cite{Ehehalt,Geiss,Cass99} for
our study of p+A, d+A and A+A collisions. This approach takes
into account the formation and
multiple rescattering of formed hadrons as well as unformed
'leading' pre--hadrons and thus incorporates the dominant final state
interactions. In the transport approach
we include nucleons, $\Delta$'s, N$^*$(1440), N$^*$(1535),
$\Lambda$, $\Sigma$ and $\Sigma^*$ hyperons, $\Xi$'s, $\Xi^*$'s
and $\Omega$'s as well as their antiparticles  on the baryonic
side whereas the $0^-$ and $1^-$ octet states are included in the
mesonic sector. Inelastic hadron--hadron collisions with energies
above $\sqrt s\simeq 2.6\GeV$ are described by the \Fritiof{}
model \cite{LUND} (employing \Pythia{} v5.5 with \Jetset{} v7.3
for the production and fragmentation of jets \cite{PYTHIA0})
whereas low energy hadron--hadron collisions are modeled in line
with experimental cross sections. We mention that no explicit
parton cascading is involved in our transport calculations
contrary to e.g.~the AMPT model \cite{Ko_AMPT} or explicit parton
cascades \cite{partons}.

A systematic analysis of HSD results and experimental data for
central nucleus--nucleus collisions for 2 AGeV to 21.3 ATeV
 has shown that the spectra for the 'longitudinal'
rapidity distribution of protons, pions, kaons, antikaons and
hyperons are in reasonable agreement with available data. Only the
pion rapidity spectra are slightly overestimated from AGS to SPS
energies \cite{Weber02,Brat03,Brat04} which implies, that the
maximum in the $K^+/\pi^+$ ratio at $20\cdots$30 AGeV -- seen in
central Au+Au (Pb+Pb) collisions \cite{NA49} -- is missed.
Furthermore, there are more severe problems with the dynamics in
the direction transverse to the beam. Whereas the pion transverse
momentum spectra are rather well described from lower AGS to top
RHIC energies the transverse momentum slopes of kaons/antikaons
are clearly underestimated above $\sim$ 5 AGeV in central Au+Au
collisions. In Ref. \cite{Brat04} this failure has been attributed
to a lack of pressure generation in the very early phase of the
heavy-ion collisions which also points towards a new state of very
strongly interacting matter in the very early phase of central
Au+Au (or Pb+Pb) collisions. In addition, the elliptic flow of
high $\pT$ hadrons is missed substantially \cite{CGG} which also
suggests strongly interacting matter in the very early phase.

Inspite of the deficiences pointed out the overall reproduction of
the experimental spectra is sufficiently realistic such that we
can proceed with more exclusive probes that are produced and
propagated in the background of the expanding string/hadron matter
generated in relativistic nucleus-nucleus collisions.

\subsection{Perturbative treatment of high $\pT$ hadrons and
jets}
For the production and propagation of hadrons with high
transverse momentum ($> 1.5\GeVc$) we employ a perturbative scheme
as also used in Refs.~\cite{CassKo,Cass01} for the charm and open
charm degrees of freedom and in Ref. \cite{CGG} for high $\pT$
hadrons.

The initial conditions for the production and also subsequent
propagation of hadrons with moderate to high transverse momentum
($>1.5\GeVc$) are incorporated in the HSD approach by a
superposition of $pp$ collisions described via PYTHIA
\cite{PYTHIA} which serve as the basic input. The partons (from
PYTHIA)  fragment to hadrons -- according to the JETSET-part of
PYTHIA -- with the vacuum fragmentation functions. We mention that
medium-modified fragmentation schemes might have to be employed
for nucleus-nucleus collisions, however, we will only use the
vacuum fragmentation functions in this study in order to find out,
if this minimal scheme also applies for the heavy-ion case.
Together with the 4-momenta of all resulting 'particles' we store
the constituent quark content and the relative weight $W_i$ of
each event.

The weight $W_i$ is given by the ratio of the actual production
cross section divided by the inelastic nucleon--nucleon cross
section, e.g.
\begin{equation}
  W_i = \frac{\sigma_{NN \rightarrow h(\pT) +
      x}(\sqrt{s})}{\sigma_{NN}^{\rm inelas.}(\sqrt{s})}.
\end{equation}
All 'particles' with one or more constituent quarks are denoted as
'pre-hadrons' while those without are denoted as 'secondary'
hadrons. The pre-hadrons start propagating in space-time  from the
space-time position of the individual interaction vertex $x^i_V$,
which is well defined from the transport calculation, and begin to
interact immediately with a fractional cross section according to
the constituent quark model (see below). The 'secondaries' are
inserted in the dynamical calculation after their formation time
$t_f$ (in the actual reference frame) at a space-time position
which is displayed from the vertex $x^i_V$ by ${\bf v}_j \cdot
t_f$ with ${\bf v}_j$ denoting the velocity of the 'secondary'
hadron $j$. The velocity ${\bf v}_j$  is fixed in the calculation
by the momentum and energy of the finally formed hadron $j$, which
is an approximation. In practice the major fraction of
'secondaries' shows up with velocities close to 1 (in units of
$c$) such that alternative modelings have no major effect on the
final results for attenuation.

We then follow the motion of the perturbative hadrons within the
full background of strings/hadrons by propagating them as free
particles, i.e.~neglecting in--medium potentials, but compute
their collisional history with baryons and mesons or quarks and
diquarks. For reactions with diquarks we use the corresponding
reaction cross section with baryons multiplied by a factor of 2/3.
For collisions with quarks (antiquarks) we adopt half of the cross
section for collisions with mesons and for the leading pre-hadron
(formed) baryon collision a factor of 1/3 is assumed. The elastic
and inelastic interactions with their fractional cross section are
modeled in the HSD approach in the same way as for ordinary
hadrons with the same quantum numbers via the \Fritiof{} model
\cite{LUND} (including \Pythia{} v5.5 with \Jetset{} v7.3 for the
production and fragmentation of jets \cite{PYTHIA0}).
 This concept -- oriented along
the additive quark model -- has been proven to work rather well
for nucleus--nucleus collisions from AGS to RHIC energies
\cite{Weber02,Brat03,Brat04} as well as in hadron formation and
attenuation in deep inelastic lepton scattering off  nuclei
\cite{Falter2,Falter}. We stress again that the latter reactions
are important to understand the attenuation of pre--hadrons or
ordinary hadrons with high momentum in 'cold' nuclear matter
\cite{Kopel4}. Our studies in Ref.~\cite{Falter2,Falter} have
demonstrated that the dominant final state interactions of the
hadrons with maximum momentum, as measured by the HERMES
collaboration \cite{HERMES}, are compatible to the concepts
described above. This also holds for antiproton production and
attenuation in proton--nucleus collisions at AGS energies
\cite{AGS02}.

\subsection{Simulation of the Cronin effect in pA, dA and AA collisions}
As known from the experimental studies of
Refs.~\cite{Cronin1,Cronin2} an enhancement of the high transverse
momentum particle cross section from proton--nucleus collisions --
relative to scaled $pp$ collisions -- is already observed at SPS
and ISR energies. This 'Cronin effect' is  probably related to an
increase of the average transverse momentum squared $\kTaveSq$ of
the partons in the nuclear medium. One may speculate that such an
enhancement of $\kTaveSq$ is due to induced initial semi--hard
gluon radiation in the medium, which is not present in the vacuum
due to  color neutrality \cite{R3,Kopel4}. Since the microscopic
mechanisms are beyond the scope of our present analysis, we
simulate this effect in the transport approach by increasing the
average $\kTaveSq$ in the string fragmentation with the number of
previous collisions $N_{\rm prev}$ as \cite{CGG,Brat04,Papp}
\begin{equation}
  \label{kt}
   \kTaveSq= \kTaveSq_{pp} (1+\alpha N_{\rm prev})\quad.
\end{equation}
The parameter $\alpha \approx 0.25 - 0.4$ in (\ref{kt}) has been
fixed in comparison to the  experimental data for d+Au collisions
at RHIC \cite{PHENIX2,STAR2} in Ref. \cite{CGG}. As discussed in
our previous analysis \cite{CGG} the Cronin effect has an
essential impact on the shape of the nuclear modification factor
\begin{equation}
  \label{ratioAA}
  R_{\rm AA}(\pT) = \frac{1/N_{\rm AA}^{\rm event}\ d^2N_{\rm AA}/dy d\pT}
  {\left<N_{\rm coll}\right>/\sigma_{pp}^{\rm inelas}\ d^2
    \sigma_{pp}/dy d\pT}\
\end{equation}
used experimentally to quantify the in-medium effects from d+Au or
A+A collisions relative to $pp$ reactions. In (\ref{ratioAA}) the
quantity $\left<N_{\rm coll}\right>$ denotes the number of binary
nucleon-nucleon collisions --  directly available from the
transport calculation -- while $\sigma_{pp}^{\rm inelas}$ stands
for the inelastic nucleon-nucleon cross section (known experimentally).

\section{Results of Transport calculations for Au+Au collisions
at $\sqrt{s}$ = 200 GeV} We recall that the detailed analysis of
high $\pT$ hadron attenuation  at the top RHIC energy of
$\sqrt{s}$ = 200 GeV in Ref. \cite{CGG} has demonstrated: i) the
(pre--) hadronic final state interactions are able to
approximately reproduce the high $\pT{}$ suppression effects
observed in d+Au and Au+Au collisions; ii) the interactions of
formed hadrons after a formation time $t_f \approx$ 0.8 fm/c  are
not able to explain the attenuation observed for transverse
momenta $\pT{} \geq$ 6 GeV/c, whereas the shape of the ratio
$R_{AA}$ (\ref{ratioAA}) in transverse momentum $\pT{}$  reflects
the presence of final state interactions of formed hadrons in the
1-4 GeV/c range; iii) the attenuation by (pre--) hadronic final
state interactions  is insufficient to explain the large
suppression seen experimentally in central Au+Au collisions. Since
our transport calculations are identical with those presented in
\cite{CGG} we directly continue with the new analysis on jet
suppression.

Fig. 3 shows our results for Au+Au collisions at \SqrtS{200} for
different centrality classes with respect to the 'near-side' and
'far-side' angular correlation of charged hadrons.  We have gated
on high $\pT$ particles in the interval
($\pT^{Trig}=4\dots6\GeV/c$) and accumulated further charged
hadrons in the interval $\pT=2\GeV\dots\pT^{Trig}$ ($|y| <0.7$).
In addition to the angular correlations from Au+Au collisions we
have added the results from
 $pp$ reactions (dashed lines) in comparison to the data
from STAR for $pp$ collisions \cite{StarAngCorr} (full dots) for
each centrality bin to demonstrate both the actual statistical
uncertainty due to different event samples for the varying
centralities (dashed lines) and the actual attenuation generated
by the final state interactions in the transport calculation
(solid lines, color: green).

We find that -- when gating on high $\pT$ hadrons (in the vacuum)
-- the 'near--side' correlations are close to the 'near--side'
correlations observed for jet fragmentation in the vacuum (cf.
lower part of Table 1) for all centralities. This is in agreement
with the experimental observation \cite{STARv2,StarAngCorr}.
Furthermore, for the 'far-side' correlations we get a $\sim$60\%
reduction for the most central collisions but not an almost
complete disappearance of the 'far-side' jet as indicated by the
experimental data \cite{STARv2,StarAngCorr}. The latter data for
central Au+Au collisions are displayed in the upper left part of
Fig. 3 by the open squares. The suppression for 5--20 \%, 20--45\%
and 45--65\% centralities in the 'far-side' peak is lower than for
the 5\% most central reactions, but not dramatically different.

%%%%%%%%%%%%%%%%%%%%%%%%%%%%%% Fig. 3 %%%%%%%%%%%%%%%%%%%%%%%%%%%%%%%%%%%
\begin{figure}[htb!]
  \begin{center}
    \includegraphics[width=13.5cm]{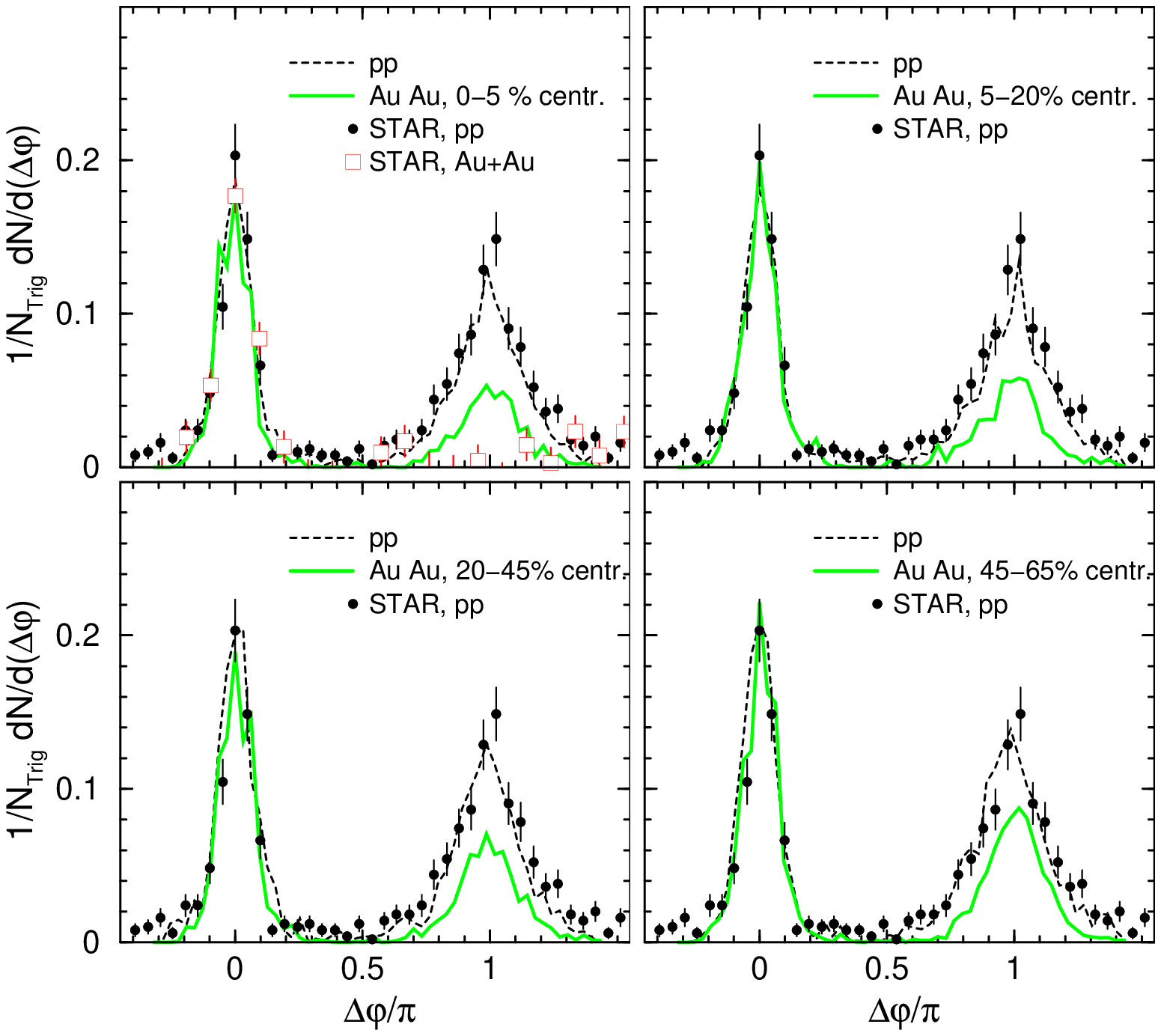}

    \caption{
      'Near-side' and 'far-side' jet-like correlations from HSD for p+p and central
Au+Au collisions at midrapidity for $\pT^{Trig}=4\dots6\GeV/c$ and
$\pT=2\GeV/c\dots\pT^{Trig}$ ($|y| <0.7$) at different
centralities. Data for p+p (full dots) are from
\protect\cite{StarAngCorr}. The dashed lines show our input to the
transport calculations for each centrality class, while the solid
lines show the result including the final state interactions
(color: green).  The experimental data
\protect\cite{STARv2,StarAngCorr} for central Au+Au collisions are
displayed in the upper left part  by the open squares (color:
red). Note, however, that the experimental trigger is
$\pT^{Trig}=4\dots6\GeV/c$ and further charged hadrons have been
accumulated in the interval $\pT=2\GeV\dots\pT^{Trig}$ for
$|\Delta \eta| <0.5$ with respect to the trigger particle. Thus
the trigger conditions differ slightly from those in the
calculations. }
    \label{angcorr}
  \end{center}
\end{figure}
%%%%%%%%%%%%%%%%%%%%%%%%%%%%%%%%%%%%%%%%%%%%%%%%%%%%%%%%%%%%%%%%%%%%%%%%%

\begin{figure}[htb!]
  \begin{center}
    \includegraphics[angle=-90,width=9cm]{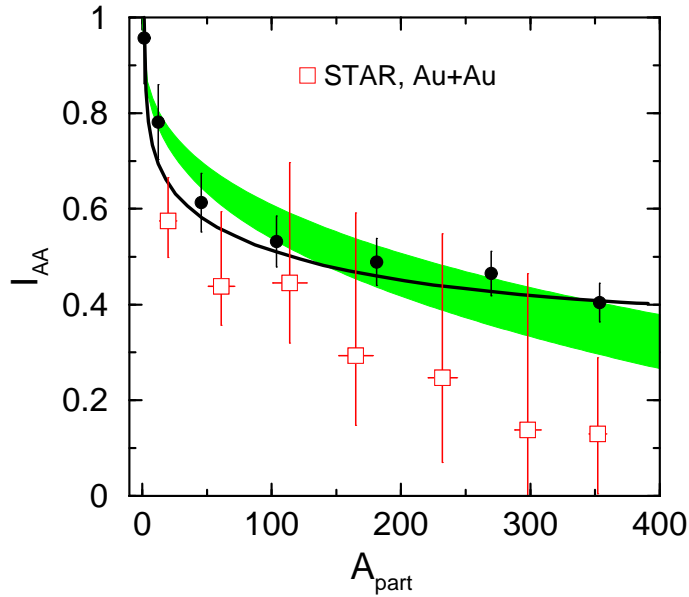}

    \caption{The angular integral over the 'far-side' correlation (\ref{intt})
    for Au+Au at $\sqrt{s}$
    = 200 GeV from HSD (full dots with statistical error bars)
    as a function of the number of participating nucleons $A_{part}$.
    The shaded area shows an attenuation $\sim A_{part}^{1/3}$, where
    the relative strength has been fixed by the calculated result either at low or high $A_{part}$,
    respectively.
    The solid line results from the simple estimate (\ref{geo}) when plotted versus
    $A_{part}$ and fixed in size at high $A_{part}$.  The experimental
data \protect\cite{STARv2} are displayed in terms of the open
squares (color: red) where the statistical and systematic errors
have been added quadratically.  Note, however, that experimentally
the associated charged hadrons have been accumulated in the
interval $\pT=2\GeV\dots\pT^{Trig}$ for $|\Delta \eta| <0.5$ with
respect to the trigger particle. Furthermore, the angular
integrations have been carried out in a slightly larger angular
range.}
    \label{fig4}
  \end{center}
\end{figure}

 In order to
become more quantitative, we show in Fig. 4 the integral over the
'far-side' peaks (cf. Fig. \ref{fig4})
\begin{equation}
\label{intt}
I_{AA} = \frac{\int_{\pi/2}^{3\pi/2}  d\varphi \ C(\Delta
\varphi)_{AA}}{\int_{\pi/2}^{3\pi/2}  d\varphi \ C(\Delta
\varphi)_{pp}}
\end{equation}
as a function of the number of participating nucleons $A_{part}$
normalized to the respective integral for $pp$ reactions. The HSD
calculations (full dots) indicate a rather fast drop with
$A_{part}$ and almost become constant for $A_{part} > $ 150. We
note that experimentally the decrease of a very similar ratio
appears to continue more strongly with $A_{part}$ \cite{STARv2}
(open squares with error bars in Fig. 4). For orientation the
shaded area (color: green) shows an attenuation $\sim
A_{part}^{1/3}$, where the relative strength has been fixed by
either the calculated result at low or high $A_{part}$,
respectively. Apparently, this power law in $A_{part}$ is too
strong in comparison to the transport calculation (full dots).

%%%%%%%%%%%%%%%%%%%%%%%%%%%%%% Fig. 4 %%%%%%%%%%%%%%%%%%%%%%%%%%%%%%%%%%%
%%%%%%%%%%%%%%%%%%%%%%%%%%%%%%%%%%%%%%%%%%%%%%%%%%%%%%%%%%%%%%%%%%%%%%%%%

We note that a rather good
reproduction of the HSD results is obtained when comparing to the
quadratic average of the long and short axis of the almond shaped
interaction zone -- in $x,y$ direction -- as a function of impact parameter $b$,
\begin{equation}
\label{geo} L(b)= \frac{1}{2} \sqrt{R_L^2+R_S^2}
\end{equation}
with $R_L=\sqrt{R^2-b^2/4}$, $R_S=R-b/2$ and $R$ denoting the
radius of the Au-target in hard-sphere geometry. The resulting
suppression is displayed in terms of the solid line in Fig. 4 --
fixed in absolute magnitude at high $A_{part}$ -- using the
relation $A_{part}(b)$. In view of the agreement between the rough
estimate  (\ref{geo}) and the numerical results of the HSD
calculations we conclude that the suppression in the transport
calculations is essentially determined by a linear geometrical
length scale, i.e. by the average jet propagation length in the
almond shaped reaction zone (cf. Ref. \cite{Drees}). Note in
addition that the relation $A_{part}(b)$ is model dependent for
very peripheral reactions (small $A_{part}$).

Since the suppression of 'far-side' jets from pre-hadronic final
state interactions in our transport calculations is less than 60\%
at all centralities, there should be an additional suppression
mechanism at least for mid-central and central collisions that is
not included in the present transport approach. Since the dynamics
at intermediate and late times are expected to be properly
described by HSD only the very early phase of the nucleus-nucleus
collision should be failed.

We recall that in HSD (as well as UrQMD \cite{URQMD1,URQMD2})
there are dominantly collisions of pre-hadrons (string ends) in
the early collision phase which neither generate enough early
transverse pressure \cite{Brat04}, elliptic flow
\cite{Brat03,Stoecker,bleichers} or attenuation of high $\pT$
hadrons \cite{CGG,CGG2}. These general findings are in line with
those seen in the suppression of 'far-side' jets in Figs. 3 and 4:
the medium of strings (pre-hadrons) is not interacting strongly
enough to explain an almost  complete quenching of 'far-side' jets
in central Au+Au collisions at $\sqrt{s}$ = 200 GeV. Nevertheless,
about 60\% (in central Au+Au collisions) is still a considerable
amount of attenuation that should be missed in parton scenarios
based on perturbative QCD cross sections.

%%%%%%%%%%%%%%%%%%%%%%%%%%%%%% Fig. 5 %%%%%%%%%%%%%%%%%%%%%%%%%%%%%%%%%%%
\begin{figure}[htb!]
  \begin{center}
    \includegraphics[width=12.5cm]{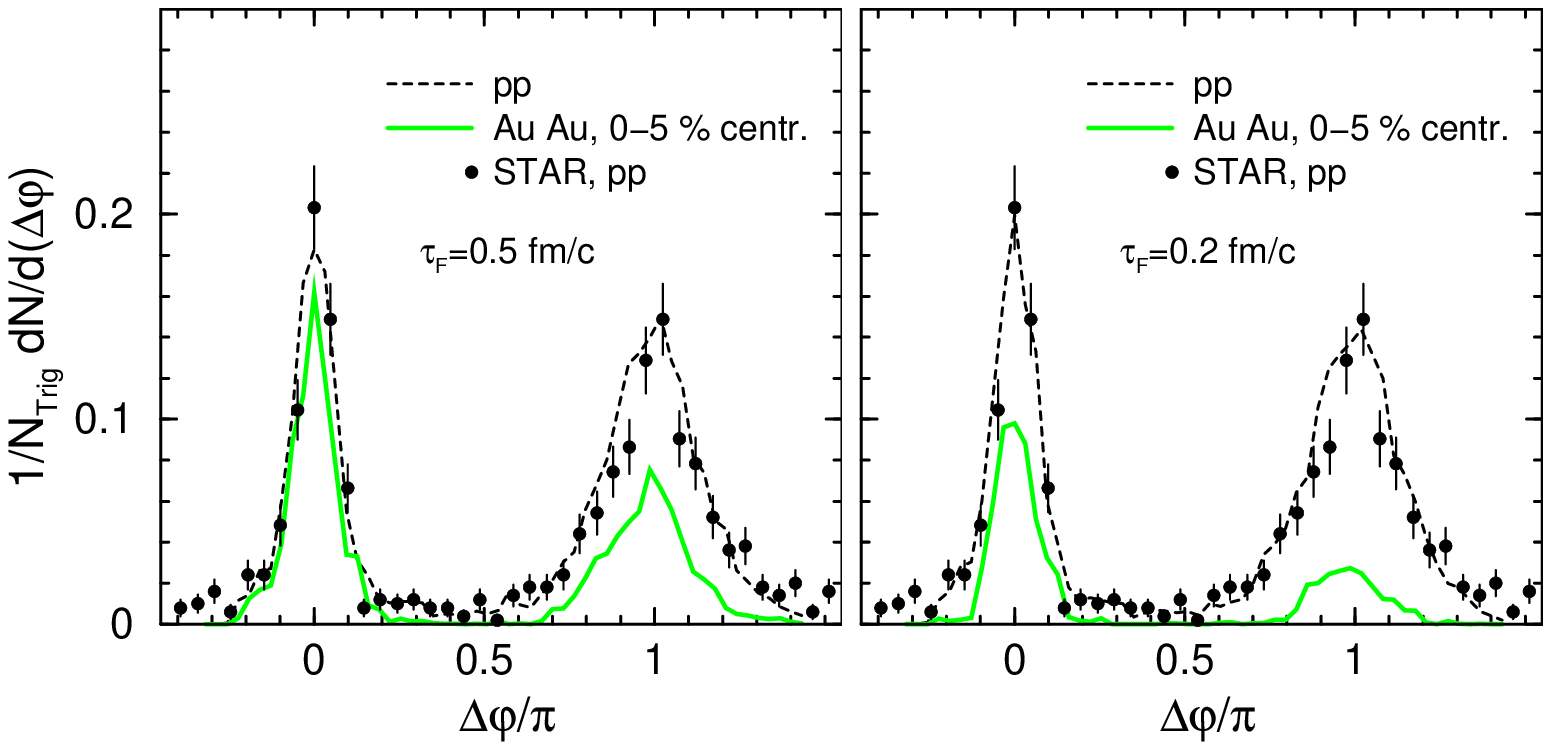}

    \caption{The same as in Fig.~\ref{angcorr} for most central collisions (upper left part),
    but now for formation times $\tau_f$ = 0.5 fm/c (l.h.s.) and 0.2 fm/c (r.h.s.)
    instead of the default value of 0.8 fm/c in Fig.~\ref{angcorr}.}
    \label{fig5}
  \end{center}
\end{figure}
%%%%%%%%%%%%%%%%%%%%%%%%%%%%%%%%%%%%%%%%%%%%%%%%%%%%%%%%%%%%%%%%%%%%%%%%%

As mentioned before the hadron formation time $\tau_f$ is a
sensible parameter of the transport approach although fixed by
hadron rapidity distributions \cite{Ehehalt}. Nevertheless, one
might argue that this formation time is simply not short enough
and an alternative 'fixing' might yield the proper jet
suppression. A similar analysis e.g. has been performed in Ref.
\cite{bleichers} with respect to the strength of the elliptic flow
$v_2$ of charged hadrons, where shorter formation times $\tau_f$
lead to an increase of the interaction rate at short times and
consequently to a rise of the elliptic flow.  To this aim we have
performed calculations for 5\% central Au+Au collisions at
$\sqrt{s}$ = 200 GeV introducing an independent formation time
$\tau_f$ for the jets and their fragmentation products without
changing the formation time of the bulk 'non-perturbative'
hadrons. The results of these calculations for the azimuthal
angular correlations are shown in Fig. 5 for $\tau_f$ = 0.5
(l.h.s.) and 0.2 fm/c (r.h.s.) by the solid lines (color: green).
In fact, the attenuation of the 'far-side' peak increases with
decreasing formation time. However, also the 'near-side' peak now
shows a reduction by the final state interactions as seen from a
comparison of the dashed and solid lines. This reduction of the
'near-side' peak is still very modest for $\tau_f$ = 0.5 fm/c but
reaches a level of $\sim$ 50\% for $\tau_f$ = 0.2 fm/c which can
clearly be ruled out by the present data. In short: a sizeable
reduction of the formation time - though leading to a more
substantial suppression of the 'far-side' peak as well as to an
increase of the elliptic flow of charged hadrons
\cite{Humanic,bleichers} -- is incompatible with the experimental
observations for central Au+Au collisions.

Without explicit representation we mention that a decrease in the
energy density cut for hadron formation (cf. Ref. \cite{CGG}) from
1 GeV/fm$^3$ to 0.7 GeV/fm$^3$ leads only to a small reduction of
the 'far-side' jet suppression. This also holds for a moderate
increase of the cross section for the pre-hadronic states.  These
results are easy to understand: The 'leading' pre-hadrons in
central collisions are attenuated by a factor of about 15 (cf.
Fig. 8 of \cite{CGG}) with the standard cross sections, such that
these particles are practically erased. Furthermore, only 70-80\%
of high $\pT$ hadrons are pre-hadrons (cf. Fig. 3 in \cite{CGG}).
We point out that a larger cross section would violate unitarity
since the full hadronic cross section for short time scales is
already exploited within the constituent quark model. Accordingly,
the 'secondaries' have to start with a vanishing cross section at
the space-time point of the vertex or more precisely starting from
the quark-antiquark production point that we denote by $t_p$ (cf.
Fig. 3 in \cite{Falter2}) which is a fraction of 1 fm/c.  During
the early times, when the system is very dense, the 'secondaries'
thus have small or vanishing cross section  up to their hadron
formation time $t_f$ (1). The latter becomes large for high $\pT$
momenta such that the attenuation of 'secondaries' is rather low
when employing high $\pT$ cuts.

%%%%%%%%%%%%%%%%%%%%%%%%%%%%%% Fig. 6 %%%%%%%%%%%%%%%%%%%%%%%%%%%%%%%%%%%
\begin{figure}[htb!]
  \begin{center}
    \includegraphics[width=11.5cm]{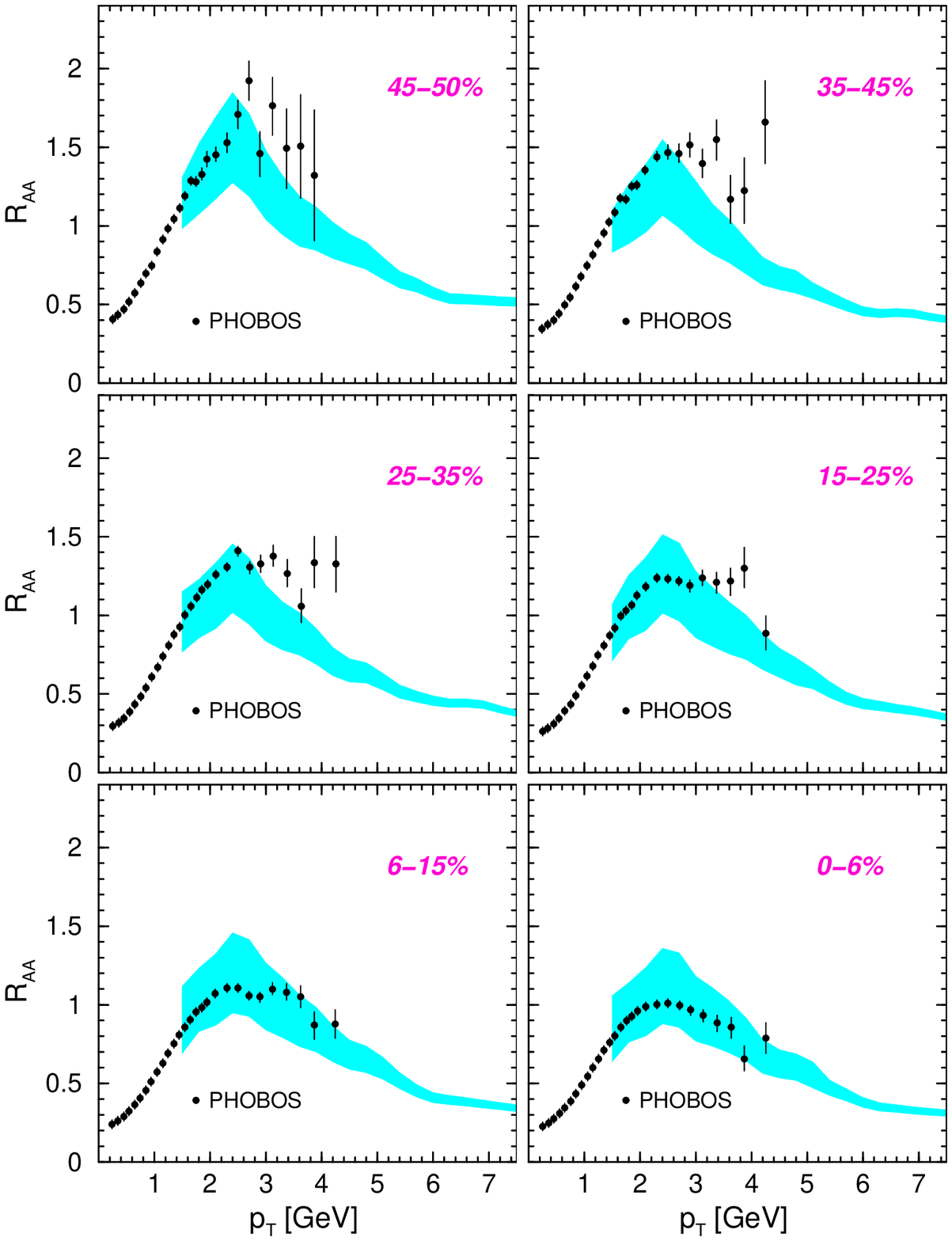}
    \caption{
      The ratio $R_{\rm AA}$ (\ref{ratioAA}) of charged hadrons for different centralities
       for Au+Au at $\sqrt{s}$ = 62.4 GeV.        The experimental data are from
      Ref.~\protect\cite{PHOBOSnew}. The hatched bands correspond
      to our calculations with a Cronin
      parameter $\alpha$ in (\ref{kt}) ranging from 0.25 to 0.4 as in Ref. \cite{CGG}. }
    \label{fig11}
  \end{center}
\end{figure}
%%%%%%%%%%%%%%%%%%%%%%%%%%%%%%%%%%%%%%%%%%%%%%%%%%%%%%%%%%%%%%%%%%%%%%%%%

\section{Au+Au collisions at $\sqrt{s}$ = 62.4 GeV}

The system Au+Au at $\sqrt{s}$ = 62.4 GeV has recently been
investigated experimentally at RHIC and first results on hadron
attenuation have become available at moderate $\pT$ from the
PHOBOS Collaboration \cite{PHOBOSnew}. We thus start with single
hadron attenuation in the next subsection following exactly the
studies in Ref. \cite{CGG} that have been performed at $\sqrt{s}$
= 200 GeV. Since all details have been presented in the former
publication we step on with the actual results. However, it is
worth to point out that the 'Cronin enhancement factors' are
higher at $\sqrt{s}$ = 62.4 GeV than at $\sqrt{s}$ = 200 GeV due
to steeper hadron spectra in $\pT$ (cf. also Ref. \cite{Dainese}).

\subsection{High $\pT$ hadron suppression}

The centrality dependence of the ratio $R_{\rm AA}$ for charged
hadrons is shown in Fig. 6 for $0\proz{}\cdots6\proz{}$,
$6\proz{}\cdots15\proz{}$,$15\proz{}\cdots25\proz{}$,
$25\proz{}\cdots35\proz{}$, $35\proz{}\cdots45\proz{}$ and
$45\proz{}\cdots50\proz{}$ centrality of Au+Au collisions at
\SqrtS{62}. Again the hatched bands (color: light blue) correspond
to our calculations with a Cronin parameter $\alpha$ in (\ref{kt})
ranging from 0.25 to 0.4 while the data stem from
Ref.~\cite{PHOBOSnew}. Note, that the uncertainty in the Cronin
enhancement (width of the hatched band) decreases slightly for
more peripheral reactions, which is a direct consequence of the
lower number of hard $NN$ collisions in (\ref{kt}).

We emphasize, that the Cronin enhancement is most visible at
momenta up to 5 GeV/c but is practically negligible for $\pT >$ 6-8
GeV/c. As pointed out before in Ref. \cite{CGG}, the suppression
seen in the calculation for larger transverse momentum hadrons is
due to the interactions of the leading (pre-)hadrons with
target/projectile nucleons and the bulk of low momentum hadrons.
Accordingly, the ratio $R_{\rm AA}$ drops below 1 in Fig. 6 in the
high momentum regime reaching an attenuation which is about the same as
 that in the calculations for $\sqrt{s}$ = 200 GeV and $\pT > $ 5 GeV/c.
Our numerical results in Fig. 6 are roughly compatible with the
data from the PHOBOS Collaboration \cite{PHOBOSnew} but
measurements at higher transverse momenta will be necessary to
prove/disprove the dropping of the ratio $R_{\rm AA}$ with
momentum $\pT$ (cf. also Ref. \cite{Miklosnew}).

%%%%%%%%%%%%%%%%%%%%%%%%%%%%%% Fig. 7 %%%%%%%%%%%%%%%%%%%%%%%%%%%%%%%%%%%
\begin{figure}[htb!]
  \begin{center}
    \includegraphics[angle=-90,width=8cm]{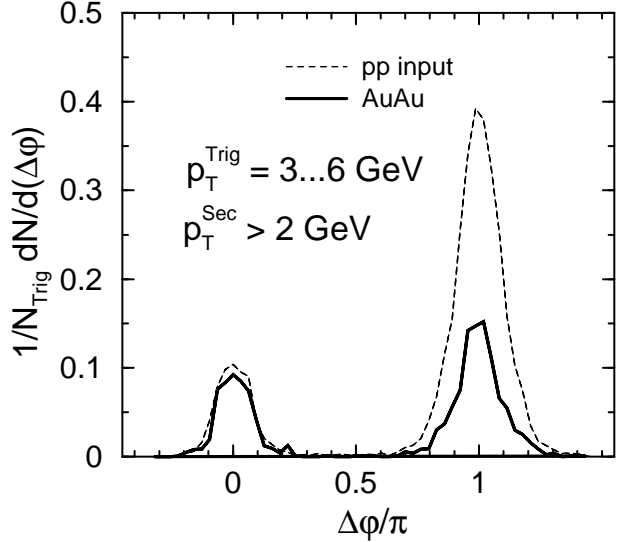}

    \caption{
      'Near-side' and 'far-side' jet-like correlations from HSD for
p+p (dashed lines) and central Au+Au collisions (solid lines) at
midrapidity for $\pT^{Trig}=3\dots6\GeV/c$ with
$\pT=2\GeV/c\dots\pT^{Trig}$
 at $\sqrt{s}$ = 62.4 GeV.
}
    \label{fig7}
  \end{center}
\end{figure}
%%%%%%%%%%%%%%%%%%%%%%%%%%%%%%%%%%%%%%%%%%%%%%%%%%%%%%%%%%%%%%%%%%%%%%%%%

\subsection{Jet correlations from $pp$ and Au+Au collisions}
Whereas first experimental information on the single hadron
attenuation is available (cf. previous subsection) the quenching
of 'far-side' jets at $\sqrt{s}$ = 62.4 GeV is still unknown from
the experimental side. In this respect we provide predictions on
the basis of our HSD calculations for the angular correlations in
$pp$ and Au+Au collisions.

In Fig. 7 we show the angular correlations of 'near-side' and
'far-side' charged hadrons by gating on high $\pT$ particles in
the interval ($\pT^{Trig}$ = $3\dots6\GeV/c$) and accumulating
further charged hadrons in the interval $\pT=2\GeV\dots\pT^{Trig}$
($|y| <1.0$). In addition to the angular correlations from central
Au+Au collisions (solid lines) we have added the results from $pp$
reactions (dashed lines) as resulting from the PYTHIA
calculations. These PYTHIA results serve again as reference lines
for the angular correlation without any final state interaction.

It is seen from the $pp$ as well as Au+Au angular correlations
that the 'far-side' peaks are more pronounced (within the cuts
taken) than the 'near-side' peaks. This effect can be traced back
to the low probability of observing a further hadron close to the
'near-side' peak when gating on a high momentum trigger particle;
most of the associated hadrons can be found in the 'far-side'
peak. This comes about because the trigger particle is not counted
in the 'near-side' correlation whereas the 'far-side' correlation
includes the leading particle of the partner jet. This explanation
is quantified in Table 2 where the probability for having 0$\dots$
3  further charged particles in the 'near-side' and 'far-side'
region is given for $pp$ (upper part) and central Au+Au collisions
at $\sqrt{s}$ = 62.4 GeV.
%%%%%%%%%%%%%%%%%%%%%%%%%%%%%%%%%%%%%% Table 3 %%%%%%%%%%%%%%%%%%%%%%%%%
\begin{table}[htb!]
  \begin{center}
\begin{tabular}{|c||cccc|}
\hline \#particles & \multicolumn{4}{c|}{\#particles far side}\\
near side      &    0      &      1     &      2     &     3
\\ \hline \multicolumn{5}{|c|}{pp @ 62.4 GeV}\\ \hline
    0 & 0.82 &  0.13 &  0.014 &  0.00062  \\
    1 & 0.026 &  0.011 &  0.0032 &  0.00029  \\
    2 & 0.00068 &  0.00061 &  0.00017 &  0.000041  \\
    3 & 0.000022 &  0.000007 &  0.000023 &  0.0000028  \\
\hline
\multicolumn{5}{|c|}{central Au+Au @ 62 GeV}\\
\hline
    0 &  0.91 &  0.051 &  0.0036 &  0.00011 \\
    1 &  0.026 &  0.0038 &  0.00061 &  0.000014 \\
    2 &  0.00033 &  0.000017 &  0.000013 &  0 \\
    3 &  0 &  0 &  0 &  0 \\
\hline
\end{tabular}
\caption{The conditional probability to find in addition to the
trigger particle 0,1,\dots3 particles at the far side and
0,1,\dots3 particles at the near side in p+p and central Au+Au
collisions at \SqrtS{62}. The trigger conditions are ($\pT^{Trig}$
= $3\dots6\GeV/c$) and $\pT=2\GeV\dots\pT^{Trig}$ ($|y| <1.0$) for
the associated hadrons.} \label{tab3}
  \end{center}
\end{table}
%%%%%%%%%%%%%%%%%%%%%%%%%%%%%%%%%%%%%%%%%%%%%%%%%%%%%%%%%%%%%%%%%%%%%%%%

The relative suppression of the 'near-side' peak is very moderate
whereas the relative suppression of the 'far-side' correlation in
central Au+Au relative to $pp$ reactions is $\sim$ 60 \% as for
$\sqrt{s}$ = 200 GeV. This is basically due to the fact that the
cross section of pre-hadrons with ordinary hadrons or
quark-diquark fragments of the initial hadrons is almost
independent on energy and the string density only moderately lower
at $\sqrt{s}$ = 62.4 GeV. On the other hand, ordinary hadrons are
allowed to form slightly earlier because the energy density (in
the local rest frame) drops faster below 1 GeV/fm$^3$.

\section{Summary}

Summarizing, we point out that (pre-) hadronic final state
interactions contribute significantly to the high $\pT$
suppression effects observed in Au+Au collisions at RHIC. This
finding is important, since the same dynamics also describe the
hadron formation and attenuation in deep--inelastic lepton
scattering off nuclei at HERMES \cite{Falter2,Falter} appreciably
well. In particular, it has been demonstrated that the centrality
dependence of the modification factor $R_{\rm AA}$ (\ref{ratioAA})
in Au+Au collisions at $\sqrt{s}$ = 200 GeV is well described for
peripheral and mid--central collisions on the basis of leading
pre-hadron interactions \cite{CGG} whereas the attenuation in
central Au+Au collisions is noticeably underestimated. A similar
observation also holds for an analysis of the 'far-side' angular
correlations, which show a substantial, but not fully complete
suppression in central collisions as indicated by the experimental
data \cite{STARv2}. From these results one should conclude that
there are some additional (and earlier) interactions of partons in
a possibly colored medium that have not been accounted for in our
present HSD transport studies.

We have, furthermore, explored the possibility if shorter
formations times $\tau_f$ for the hadrons might lead to an
improved description of the angular correlation data since smaller
$\tau_f$ lead to an increase of the interaction rate at early
times and to an increase of the elliptic flow of charged hadrons
\cite{Humanic,bleichers}. However, for formation times $\tau_f
\leq$ 0.5 fm/c also the 'near-side' peak shows a reduction by the
final state interactions.  This reduction of the 'near-side' peak
is still very modest for $\tau_f$ = 0.5 fm/c but reaches a level
of $\sim$ 50\% for $\tau_f$ = 0.2 fm/c which can be ruled out by
the present data. In short: a sizeable reduction of the formation
time - though leading to a more substantial suppression of the
'far-side' peak and to an increase of the elliptic flow $v_2$ of
hadrons -- is incompatible with the experimental observations for
central Au+Au collisions.

We close in pointing out that further experimental studies on the
suppression of high momentum hadrons from d+Au and Au+Au
collisions down to $\sqrt{s}$ = 20 GeV will be necessary to
clearly separate initial state Cronin effects from final state
attenuation and to disentangle the role of partons in a colored
partonic medium from those of  interacting pre-hadrons in an
approximately color-neutral hot and dense fireball. A major
problem, however, is that the jets escaping  the medium stem from
the corona of the fireball \cite{Dainese,Eskola,Drees,MuellerB}
and thus do not carry much information from the dense medium.
Nevertheless, 'far-side' jet suppression appears to be a promising
observable since the pre-hadronic (color neutral) interactions
employed so far lead to a suppression that is dominantly
characterized by a linear length scale (cf. the comparison in Fig.
4). Thus when gating on peripheral reactions with $A_{part}$ from
50 to 150 one might exploit the geometry of the fireball -- which
is of approximate almond shape -- in more detail. To this aim
narrow bins in $A_{part}$ should be performed and the di-hadron
correlations be investigated as a function of the azimuthal angle
$\phi$ of the trigger hadron relative to the reaction plane.
Furthermore, also several cuts in the minimum transverse momentum
of the associated particles should be taken. It is clear that very
high statistics for jet correlations will be needed to obtain a
solid information. Some experimental work along this direction is
presently carried out.

\vspace{0.5cm}
The authors like to acknowledge stimulating discussions with
C. Greiner, T. Falter, B. Kopeliovich and I. Vitev throughout this study.

\end{document}